\documentclass[
    aps,
    prl,
    twocolumn,
    superscriptaddress,
    amsfonts,
    amssymb,
    amsmath,
    eqsecnum
]{revtex4-2}

\usepackage{xcolor}
\definecolor{darkred}{RGB}{220, 0, 0}

\definecolor{bluer}{RGB}{50, 50, 150}
\definecolor{redr}{RGB}{220, 50, 50}
\definecolor{pinkr}{RGB}{200, 0, 100}

\usepackage{braket}
\usepackage{bm}
\usepackage{mathtools}
\usepackage{dsfont}
\usepackage{mathrsfs}
\usepackage{graphicx}
\usepackage[
    colorlinks=true,
    citecolor=bluer,
    linkcolor=redr,
    urlcolor=pinkr,
]{hyperref}
\usepackage{cleveref}
\usepackage[shortlabels]{enumitem}
\usepackage{nccmath}
\usepackage{accents}

\DeclarePairedDelimiterXPP\tmpTr[1]{\mathrm{Tr}}{[}{]}{}{#1}
\newcommand{\Tr}{\tmpTr*}
\DeclarePairedDelimiterXPP\tmpE[1]{\mathbb{E}}{[}{]}{}{#1}
\newcommand{\E}{\tmpE*}
\DeclarePairedDelimiterXPP\tmppr[1]{\mathbb{P}}{[}{]}{}{#1}

\newcommand{\dd}{\mathrm{d}}
\newcommand{\dt}{\dd t}
\newcommand{\dYt}{\dd Y_t}

\newcommand{\ddp}{\dd p}
\newcommand{\dx}{\dd x}
\newcommand{\Dt}{\Delta t}
\newcommand{\sDt}{\sqrt{\Dt}}
\newcommand{\brho}{\bm{\rho}}
\newcommand{\dbY}{\dd\bm{Y}}
\newcommand{\bI}{\bm{I}}

\newcommand{\knowing}{\,|\,}

\newcommand{\bY}{\bm{Y}}
\newcommand{\bW}{\bm{W}}

\newcommand{\lbb}{\{\hspace{-0.38em}\{}
\newcommand{\rbb}{\}\hspace{-0.38em}\}}

\crefname{figure}{figure}{figures}
\Crefname{figure}{Figure}{Figures}

\newcommand{\robinet}{\overline{\rho}}

\newcommand{\Ccal}{\mathcal{C}}
\newcommand{\Kcal}{\mathcal{K}}
\newcommand{\Lcal}{\mathcal{L}}
\newcommand{\Ical}{\mathcal{I}}
\newcommand{\Hrm}{\mathrm{H}}
\newcommand{\Irm}{\underaccent{\bar}{I}}

\begin{document}

\title{Time-averaged continuous quantum measurement}

\author{Pierre Guilmin}
\email{pierre.guilmin@alice-bob.com}
\affiliation{Alice \& Bob, 49 Bd du Général Martial Valin, 75015 Paris, France}
\affiliation{Laboratoire de Physique de l’École Normale Supérieure, Mines Paris, Inria, CNRS, ENS-PSL, Centre Automatique et Systèmes (CAS), Sorbonne Université, PSL Research University, Paris, France}

\author{Pierre Rouchon}
\affiliation{Laboratoire de Physique de l’École Normale Supérieure, Mines Paris, Inria, CNRS, ENS-PSL, Centre Automatique et Systèmes (CAS), Sorbonne Université, PSL Research University, Paris, France}

\author{Antoine Tilloy}
\affiliation{Laboratoire de Physique de l’École Normale Supérieure, Mines Paris, Inria, CNRS, ENS-PSL, Centre Automatique et Systèmes (CAS), Sorbonne Université, PSL Research University, Paris, France}

\begin{abstract}
    \noindent The theory of continuous quantum measurement allows to reconstruct the state $\rho_t$ of a system from a continuous stochastic measurement record $I_t$. However, this truly continuous-time signal $I_t$ is \emph{never} available in practice. In experiments, one generally has access to its digitization, i.e., to a series of time averages $I_k$ over finite intervals of duration $\Delta t$. In this letter, we take this digitization seriously and define $\bar{\rho}_n$ as the best Bayesian estimate of the quantum state given (only) a digitized record $(I_1,\dots,I_n)$. We show that $\bar{\rho}_{n+1}$ can be computed recursively from $I_{n+1}$ and $\bar{\rho}_n$ using an exact formula. The latter can be evaluated numerically exactly, or used as the basis for a perturbative expansion into successive powers of $\sqrt{\Delta t}$. This allows reconstructing quantum trajectories in regimes of coarse $\Delta t$ where existing methods fail, estimating parameters at fixed $\Delta t$ without bias, and directly sampling digitized quantum trajectories with schemes of arbitrarily high order.
\end{abstract}

\maketitle

\noindent \textbf{Introduction} -- The continuous measurement of quantum systems has become commonplace across various experimental platforms, from quantum optics to superconducting circuits. These experiments are modeled with the formalism of stochastic master equations (SME), which describes how the state of the system $\rho_t$ at time $t$ depends on the continuously measured signal $\{I_s\}_{s\leq t}$ up to that time \cite{wiseman2009quantum}. However, the observer never has access to this signal (which would require infinite memory to store) but only to its digitized version, sampled at regular time intervals $\Dt$. The only observable quantity is thus the discrete-time signal $(I_1,\dots, I_n)$, where each $I_k$ is defined as the time average of the continuous-time signal:
\begin{equation}\label{eq:binned_signal_def}
    I_k = \int_{(k-1)\Dt}^{k\Dt} I_t\,\dt = \int_{(k-1)\Dt}^{k\Dt}\dYt ~\text{with}~ I_t = \frac{\dYt}{\dt}.
\end{equation}
The standard strategy is to take $\Dt$ as small as experimentally feasible with modern analog-to-digital converters, and to solve the SME by directly using the digitized signal $I_k$ \emph{as if} it were the continuous signal $I_t$.

This approximation only holds when $\Dt$ is infinitely smaller than all other dynamical timescales. This is not always the case for modern experimental platforms, especially with superconducting circuits. This is a problem in principle, as it is unclear how to refine the approximation if needed, but also in practice, as it can lead to incorrect state reconstruction or biased parameter estimation. More insidiously, this naive strategy ties together two \emph{a priori} orthogonal constraints on $\Dt$: it must be small enough to capture most of the information available in the continuous signal, but also small enough to ensure that reconstructing the state with the discretized SME is numerically stable. Only the former constraint is physically relevant, but the latter is often the bottleneck, forcing experimentalists to store and process far more data than necessary.

To improve upon this, the first step is to introduce the correct object. This requires an important if somehow trivial observation: once the signal has been averaged over a finite duration, some information is inevitably lost. Thus, there is no hope to reconstruct the ``true'' continuous-time state trajectory $\{\rho_s\}_{s\leq t}$ from the measured data, because different trajectories can lead to the same value of the digitized signal. All one can ask for is the ``best'' estimate of the system state (in the Bayesian sense), obtained by averaging all the possible trajectories that correspond to the same digitized signal. Starting from a known initial state $\rho_0$ at time $t=0$, and given a digitized signal $(I_1, \dots, I_n)$, we define the state $\robinet_n$ at time $t=n \Dt$ by
\begin{equation}\label{eq:robinet}
    \robinet_n = \E{\brho_{n \Dt} \knowing I_1, \dots, I_n},
\end{equation}
where $\mathbb{E}[\,\bullet \knowing I_1,\dots,I_n]$ is the conditional average over all possible trajectories starting from $\rho_0$ whose time-averaged signal matches the measured digitized signal. We call $\robinet_n$ the \emph{robinet} state \footnote{In French, robinet means faucet / tap. The name comes from pronouncing ``binned $\rho$'' in improper French, which gives ``rho binné'', which sounds like robinet.}. This letter shows this object is more manageable than it may initially appear. We derive an exact recursive formula giving $(\robinet_1, \dots, \robinet_n)$ as a function of $(I_1, \dots, I_n)$ and explain how it can be evaluated to arbitrary precision using quadratures, or used to derive new perturbative discretization schemes valid to arbitrarily high order in $\Dt$. We show how this enables to reconstruct trajectories better than existing techniques, to estimate unknown parameters without introducing systematic bias, and to directly simulate quantum trajectories at a coarse-grained timescale.

\noindent\textbf{Standard Formalism} -- First, we recall the standard formalism of SME. In the following, we distinguish a stochastic process $\mathbf{X}_t$ (bold font) from a specific realization of the process $X_t$ (normal font). For clarity, we first discuss a single diffusive measurement, and then generalize in the appendix. The usual formalism describes the joint stochastic dynamics of the system state and the measurement record $(\brho_t,\bY_t)$ at time $t$ \cite{jacobs2006straightforward}:
\begin{align}
    \dd\brho_t &= \Lcal_t(\brho_t)\,\dt + \left(\Ccal_t(\brho_t)- \Tr{\Ccal_t(\brho_t)}\,\brho_t\right)\dd\bW_t \label{eq:drho}\\
    \dd \bY_t &= \Tr{\Ccal_t(\brho_t)}\,\dt + \dd\bW_t \label{eq:dY},
\end{align}
with $\Lcal$ the Liouvillian of the system, $\bW$ a Wiener process and ${\Ccal(\bullet) = \sqrt\eta (L \bullet + \bullet L^\dag)}$ with $0\leq\eta\leq1$ the measurement efficiency and $L$ the monitored jump operator. The Liouvillian $\Lcal$ contains the dissipator $L\rho L^\dag - \frac12L^\dag L\rho-\frac12\rho L^\dag L$ associated with $L$, and may also feature other sources of dissipation and unitary evolution.

Each new experiment results in a particular realization of the measured signal $\{Y_s\}_{s\leq t}$, which corresponds to a specific state trajectory $\{\rho_s\}_{s\leq t}$. In principle, this state trajectory can be reconstructed from the continuous-time signal $\{Y_s\}_{s\leq t}$ (often written as in the introduction $\{I_s\}_{s\leq t}$ with $I_s=\dd Y_s/\dt$), by expressing $\dd W_t$ as a function of $\dYt$ in \cref{eq:drho} using \cref{eq:dY}. However, as we argued, $\rho_t$ cannot be reconstructed from the digitized signal $I_k$ defined in \cref{eq:binned_signal_def}.

\noindent\textbf{Derivation} -- To find the exact expression for the robinet state, we start from its definition \cref{eq:robinet}:
\begin{align}\label{eq:markov}
    \robinet_n &= \E{\brho_{n \Dt} \knowing \bI_1=I_1, \dots, \bI_n=I_n}\\
    &= \frac{\E{\delta(I_n - \bI_n)\dots\delta(I_1 - \bI_1)\,\brho_{n \Dt}}}{\E{\delta(I_n - \bI_n)\dots\delta(I_1 - \bI_1)}},\label{eq:delta}
\end{align}
with $\bI_k = \int_{(k-1)\Dt}^{k\Dt}\dd \bY_t$ the random variable for the $k$-th bin digitized record. We call the numerator the \emph{linear robinet state} $\tilde\rho_n$. The denominator gives the probability density of measuring the digitized record $(I_1,\dots,I_n)$. It is the trace of $\tilde\rho_n$, which ensures that $\robinet_n$ has unit trace.

Let us focus on the numerator $\tilde\rho_n$. Using the inverse Fourier transform of the Dirac delta function ${\delta(x) = \frac{1}{2\pi} \int_{\mathbb{R}} \ddp\, e^{i p x}}$, we have:
\begin{align}
    \tilde\rho_n &= \E{\delta(I_n - \bI_n)\dots\delta(I_1 - \bI_1)\,\brho_{n \Dt}} \notag\\
    &= \E{\frac{1}{(2\pi)^n} \int_{\mathbb{R}^n} \ddp_n...\ddp_1\, e^{i \sum_k p_k (I_k - \bI_k)} \brho_{n \Dt}}\label{eq:fourier}\\
    &= \frac{1}{(2\pi)^n} \int_{\mathbb{R}^n} \ddp_n...\ddp_1\, e^{i \sum_k p_k I_k}\, \E{e^{-i \sum_k p_k \bI_k} \brho_{n \Dt}}.\notag
\end{align}
To compute the expectation, we define the piecewise constant function $j_t=-ip_k$ if $(k-1)\Dt \leq t < k\Dt$ and $j_t=0$ otherwise, and we introduce
\begin{equation}
    \varrho_t^j = \E{\exp\left(\int_0^t j_s\,\dbY_s\right) \brho_t},
\end{equation}
such that at time ${t=n \Dt}$ it is exactly the expectation we want to compute: ${\varrho_{n \Dt}^j=\E{e^{-i \sum_k p_k \bI_k} \brho_{n \Dt}}}$. Using Itô's lemma \cite{guilmin2024parameters}, we find that $\varrho_t^j$ obeys a linear ODE (ordinary differential equation) governed by the \emph{tilted Liouvillian} $\Lcal^j$:
\begin{equation}
    \frac{\dd \varrho_t^j}{\dt} = \Lcal_t^j(\varrho_t^j) \text{ with } \Lcal^j = \Lcal +j\Ccal + \frac{j^2}{2}\Ical,
\end{equation}
with $\Ical$ the identity superoperator. The ODE solution at time $t$ starting from $\rho_0$ is the time-ordered exponential of the tilted generator. For brevity, we assume that the Liouvillian is time-independent which simply gives ${\varrho_t^j = e^{t \Lcal^j}}$.

We may now dissect the evolution on each time bin up to time $t=n\Dt$, by writing $\varrho_{n \Dt}^j = \Phi_{p_n}\cdots\Phi_{p_1}(\rho_0)$ with
\begin{equation}
     \Phi_p = e^{\Dt\,\Lcal^{-ip}} = e^{-\Dt\frac{p^2}{2}}e^{\Dt(\Lcal-ip\Ccal)}.
\end{equation}
Replacing the expectation by $\varrho_{n \Dt}^j$ in \cref{eq:fourier} we have
\begin{equation}
    \tilde\rho_n = \frac{1}{(2\pi)^n} \int_{\mathbb{R}^n} \ddp_n...\ddp_1\, e^{i \sum_k p_k I_k}\, \Phi_{p_n}\cdots\Phi_{p_1}(\rho_0).
\end{equation}
By linearity of $\Phi_p$, we find that the expression of $\tilde\rho_n$ depends only on the previous linear robinet state $\tilde\rho_{n-1}$:
\begin{equation}
    \tilde\rho_n = \frac{1}{2\pi} \int_{\mathbb{R}} \ddp_n\, e^{i p_n I_n} \Phi_{p_n} (\tilde\rho_{n-1}).
\end{equation}
The final robinet state $\robinet_n$ can thus be computed by iterating this equation, and normalizing by taking the trace at each step, or at the end.

\noindent\textbf{Exact map} -- The robinet states $(\robinet_1, \dots, \robinet_n)$ are reconstructed from the digitized signal $(I_1\dots,I_n)$ by iterating:
\begin{equation}\label{eq:Kcal-it}
    \robinet_k = \frac{\Kcal_{I_k}(\robinet_{k-1})}{\Tr{\Kcal_{I_k}(\robinet_{k-1})}},
\end{equation}
starting from $\robinet_0=\rho_0$. The completely positive linear map $\Kcal_I$ is defined by
\begin{equation}\label{eq:Kcal}
    \Kcal_I(\rho) = \frac{1}{2\pi} \int_\mathbb{R} \ddp\, e^{i p I - \Dt\frac{p^2}{2}} e^{\Dt(\Lcal -ip\Ccal)}(\rho).
\end{equation}
The denominator gives the probability density $p_{\bI_k}$ of measuring the digitized signal $I_k$ at step $k$:
\begin{equation}\label{eq:Kcal-prob}
    p_{\bI_k}(I_k)=\Tr{\Kcal_{I_k}(\robinet_{k-1})}.
\end{equation}
This scheme gives the exact quantum instrument corresponding to the time averaging of a continuous measurement over a duration $\Dt$. The formula is valid for any duration $\Dt$, even arbitrarily long.

The result can be easily generalized to jump measurements, where digitization accounts for the (inevitably finite) resolution on the measured jump times. The same derivation applies by (i) conditioning on the discrete-valued digitized signal $I_n\in\mathbb{N}$ with the Kronecker (instead of Dirac) delta function in \cref{eq:delta}, (ii) using the Fourier representation ${\delta_{n,0} = \frac{1}{2\pi} \int_{-\pi}^{\pi} \ddp\, e^{i p n}}$ of the Kronecker delta in \cref{eq:fourier} and (iii) introducing the tilted Liouvillian for jump measurement ${\Lcal^j=\Lcal + (e^j-1)\, \Ccal}$ with ${\Ccal(\bullet)=\theta\bullet+\eta L\bullet L^\dag}$ (where $\theta\geq0$ is the dark count rate) \cite{rouchon2022tutorial, guilmin2024parameters}. The formula also generalizes to multiple monitored loss operators, see appendix.

\noindent\textbf{Evaluating the formula} -- We provide two ways to evaluate the linear map \cref{eq:Kcal}: one numerically exact, valid for arbitrarily large $\Dt$, and one perturbative, particularly illuminating analytically and usable for small $\Dt$.

The first option is to compute the integral over $p$ using a Gauss-Hermite quadrature:
\begin{equation}\label{eq:quadrature_equality}
    \int_{\mathbb{R}} \ddp\, e^{-p^2} f(p) \approx \sum_{j=1}^{n_p} w_j f(p_j),
\end{equation}
where $w_j$ and $p_j$ are the quadrature weights and roots. For the smooth integrands we are considering, such quadratures are numerically exact for a few dozen points. In the jump case, the Gauss-Legendre quadrature can be used instead.

For small Hilbert space dimensions, we can compute the full map $\Kcal_{I_k}$, by precomputing once the superoperators $e^{\Dt(\Lcal-ip_j\Ccal)}$ for all the quadrature points $p_j$. For large Hilbert space dimensions, it is more efficient to compute the action of $\Kcal_{I_k}$ on $\robinet_{k-1}$ without ever writing the full map explicitly. In that case, we evaluate the integrand for a specific quadrature point $p_j$ by solving the ODE with generator $\Lcal-ip_j\Ccal$ for a duration $\Dt$. The result can be computed to arbitrary numerical precision, either by using a high-order Runge-Kutta method, or, for a constant Liouvillian, by computing the action of the exponential map on the state using a Krylov subspace method. The numerical cost of computing $\Kcal_I(\rho)$ is thus the same as solving $n_p$ Lindblad-like master equations (about $n_p\approx 30$ for the examples in this paper, which can be computed in batch).

An alternative approach, when $\Dt$ is smaller than the other timescales of the dynamics, but not infinitely small, is to compute the integral by Taylor expanding in $\Dt$. For a sufficiently small time bin, the digitized signal $I$ is of order $\sDt$, so we introduce the reduced variable $\Irm=I/\sDt$ of order one. We then expand the map $\Kcal_I$ to arbitrary order in $\sqrt{\Dt}$ to obtain (see appendix):
\begin{equation}
    \Kcal_I = \frac{e^{-\Irm^2/2}}{\sqrt{2\pi\Dt}} \sum_{q=0}^\infty \sDt^q \!\!\!\sum_{n=\lceil q/2\rceil}^{q}\!\!\!\frac{\Hrm_{2n-q}(\Irm)}{n!}\lbb\Lcal^{q-n} \Ccal^{2n-q}\rbb,\label{eq:Kcal_I_p}
\end{equation}
where $\Hrm_k$ is the probabilist Hermite polynomial of degree $k$ and the symbol $\lbb A^aB^b\rbb$ denotes the sum of all permutations of $a$ symbols $A$ and $b$ symbols $B$, e.g., $\lbb A^2 B^2\rbb = AABB + ABAB + ABBA + BBAA + BABA + BAAB$. The first factor is the usual Gaussian function centered at zero with standard deviation $\sDt$.

For example, at order $4$ in $\sDt$ and ignoring the Gaussian factor:
\begin{align}
    \Kcal_I^{(4)} &\propto\,\mathcal{I}+ \sDt^1\Irm\,\Ccal+ \sDt^2\left(\Lcal+\frac{\Irm^2-1}{2}\Ccal^2\right)\notag\\
    &+\sDt^3\left(\frac{\Irm}{2}\lbb\Lcal\Ccal\rbb + \frac{\Irm^3-3\Irm}{3!}\Ccal^3\right)\label{eq:Kcal_4}\\
    &+\sDt^4\left(\frac{1}{2}\Lcal^2+\frac{\Irm^2-1}{3!}\lbb\Lcal\Ccal^2\rbb+\frac{\Irm^4-6\Irm^2+3}{4!}\Ccal^4\right),\notag
\end{align}
with ${\lbb\Lcal\Ccal\rbb=\Lcal\Ccal+\Ccal\Lcal}$ and ${\lbb\Lcal\Ccal^2\rbb=\Lcal\Ccal^2+\Ccal\Lcal\Ccal+\Ccal^2\Lcal}$. The first and second orders correspond to the usual discretization, including the Milstein correction ${(\Irm^2-1)\,\Ccal^2/2}$ which was first derived in \cite{rouchon2015efficient}. The third and fourth orders contain all the terms derived in \cite{guevara2020completely} and later \cite{wonglakhon2024completely}, as well as some new ones. This new derivation extends these results to arbitrarily high orders in $\sDt$. It also justifies these perturbative expansions as the Bayesian optimum. Interestingly, for perfect detection efficiency $\eta=1$, the expansion can be written up to order five included in the form of a Kraus map with a single Kraus operator: $\Kcal_I(\rho)= M_I\rho M_I^\dag + \mathcal{O}(\sDt^6)$ (see appendix). For pure states, the loss of information due to the averaging, which leads to degraded state purity, is thus a subtle effect that only appears at order $\Dt^3$!

The number of terms in the symbolic expansion increases exponentially with the order, but we can recursively compute their sum, resulting in a quadratic cost in the chosen order (see appendix). For sufficiently small $\Dt$, this provides an alternative to the quadrature method: a simple and explicit numerical scheme to approximate $\Kcal_I(\rho)$, that does not require an external ODE solver.

\begin{figure}[ht]
    \centering
    \includegraphics{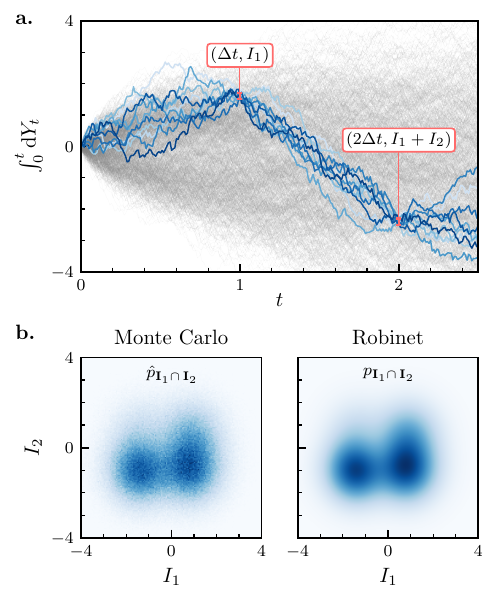}
    \caption{
        \textit{Numerical verification.} The system is a qubit starting from its excited state with Hamiltonian $H=\sigma_x+\sigma_y/2$ and a single jump operator $L=2\sigma_-$ continuously measured with efficiency $\eta=0.8$. The measured signal is digitized on two successive time bins of duration $\Dt=1.0$. \textbf{a.} Cumulative measurement $\int_0^t \dYt$ for $10^3$ trajectories (in light gray) out of $10^6$ simulated trajectories. The $10$ trajectories highlighted in blue are post-selected because they match $(I_1,I_2)=(1.5, -4.0)$ at times $(\Dt, 2\Dt)$ with precision $\epsilon=0.1$ (a $0.001\%$ post-selection rate). We verify that averaging the $10$ corresponding states at $t=2\Dt$ approximates the robinet state $\robinet_2$. \textbf{b.} Probability density of the joint random variable $(\bI_1,\bI_2)$ estimated from the $10^6$ simulated trajectories (left panel), or computed exactly as the trace of the linear robinet state $\tilde{\rho}_2$ (right panel).
    }
    \label{fig:verification}
\end{figure}

\noindent\textbf{Verifying the formula} -- We verify numerically the validity of our formula by testing it against a naive post-selection method. To this end, we first simulate a large number of fine-grained quantum trajectories. For fixed values of the digitized signal on two consecutive time bins $(I_1, I_2)$, we keep only those trajectories for which the digitized measurement $(I_1^\text{traj}, I_2^\text{traj})$ matches closely, i.e., such that $|I_1^\text{traj} - I_1|\leq\epsilon$ and $|I_2^\text{traj} - I_2|\leq\epsilon$, with $\epsilon$ a small tolerance parameter (see \cref{fig:verification}).

We average the associated states at the final time $t=2\Dt$, and verify that it matches the robinet state \cref{eq:Kcal-it} computed exactly with Gaussian quadratures. As the tolerance $\epsilon$ is decreased (at the cost of increasingly fewer post-selected states), the Monte Carlo average converges to the robinet state. To confirm the agreement, we repeated the procedure for several values of $I_1$ and $I_2$, over digitization timescales $\Dt$ spanning several orders of magnitude, and for different system dynamics.

In addition to being numerically very expensive, the Monte Carlo estimate of the robinet state is doubly inaccurate because (i) the finite number of trajectories introduces a statistical error (with variance decreasing as one over the number of trajectories), and (ii) the tolerance $\epsilon$ introduces a systematic error. In contrast, directly computing the robinet state is both numerically efficient and exact.

Another way to verify the formula is to compute exactly the probability density of the integrated signal $p_{\bI_1}(I_1)=\Tr{\Kcal_{I_1}(\rho_0)}$ or $p_{\bI_2}(I_2)=\Tr{\Kcal_{I_2}e^{\Dt \Lcal}(\rho_0)}$, or even the joint law $p_{\bI_1,\bI_2}(I_1, I_2)=\Tr{\Kcal_{I_2}\Kcal_{I_1}(\rho_0)}$, and compare with the Monte Carlo estimate from all simulated trajectories, see \cref{fig:verification}.

\noindent\textbf{Applications} -- The most immediate application of our method is for state reconstruction \cite{murch2013observing,weber2014mapping,weber2016quantum,campagne2016observing,ficheux2018dynamics}. To illustrate its power, we focus on a recent experimental setup where using the standard continuous measurement theory with digitized data proved challenging. In \cite{marquet2024autoparametric}, a peculiar dissipation between a bosonic mode housed in a superconducting resonator and its environment is engineered. The Hilbert space associated with the system is that of a harmonic oscillator with annihilation operator $a$. The (simplified) dynamics is described by the Hamiltonian $H=0$ and a single jump operator $L=\sqrt{\kappa_2}a^2$ measured with efficiency $\eta=0.2$ by continuous homodyne detection along the $X$ quadrature. Of course, additional terms can be added in practice. The dissipation rate is $\kappa_2/(2\pi)=2\,\mathrm{MHz}$ and the digitization time is $\Dt=4\,\mathrm{ns}$. We consider a ``deflate'' scenario, where one starts from an initial coherent state $\ket{\alpha=4}$ and let the system evolve under dissipation and measurement. The goal is to reconstruct the state trajectory from the digitized measurement record $(I_1,\dots,I_n)$ for $0\leq t\leq 100\,\mathrm{ns}$. We compare three reconstruction methods: (i) a naive \texttt{Euler} method, (ii) the method described in \cite{jordan2016anatomy}, which we call \texttt{CPTP-1}, and (iii) \texttt{Robinet}.

\begin{figure}[ht]
    \centering
    \includegraphics{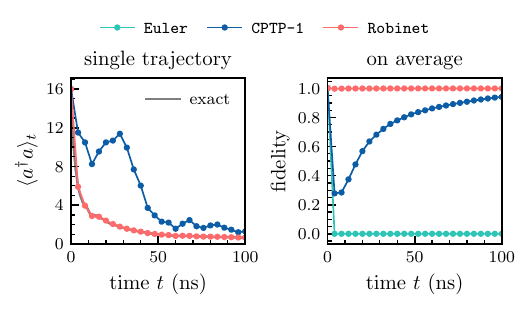}
    \caption{\textit{State reconstruction.} Predicted photon number for a single trajectory (left panel) and average fidelity for many trajectories (right panel). The state reconstructed with \texttt{Euler} is completely unphysical, so we exclude it from the left panel.}
    \label{fig:fidelity}
\end{figure}

We simulate a very fine-grained trajectory, and average the signal over the duration $\Dt$ to obtain the digitized record $(I_1,\dots,I_n)$. We run each reconstruction method with a step size $\Dt$ using this digitized signal, and compare the reconstructed states predicted photon number $\braket{a^\dag a}_t$. We then repeat this procedure for many different trajectories and compute the average fidelity between the reconstructed state and the finely simulated ``true'' state. \Cref{fig:fidelity} summarizes the results.

Since averaging results in the loss of some information, even the robinet state cannot achieve a fidelity of exactly $1$ with respect to the continuous trajectory. In this example, however, this loss of fidelity is extremely moderate. The substantial infidelity of the other schemes primarily stems from their incorrect reconstruction of the dynamics, which is not due to an insufficient amount of information in the digitized signal, but rather to their overly coarse discretization. In this example, ignoring the signal entirely and instead using the Lindblad average state would yield a better fidelity than the standard \texttt{Euler} and \texttt{CPTP-1} schemes.

\begin{figure}[ht]
    \centering\includegraphics{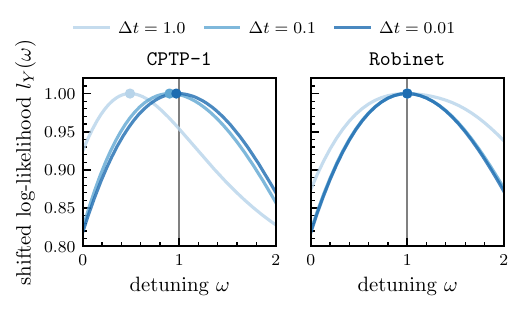}
    \caption{
        \textit{Bayesian inference.} Same example as in \cref{fig:verification} with an additional detuning term $\omega \sigma_z$ in the Hamiltonian. We simulate $10^5$ trajectories with $\omega=1.0$ and compute the log-likelihood (y-axis) for different values of $\omega$ (x-axis) and for several digitization times $\Dt$ (blue shades) with the \texttt{CPTP-1} method (left panel) or the \texttt{Robinet} method (right panel). To facilitate comparison, we shifted each log-likelihood curve vertically by a fixed amount to align its maxima (indicated by large dots).
    }
    \label{fig:maxlike}
\end{figure}

Another application of the robinet state is in the context of Bayesian parameter estimation \cite{mabuchi1996dynamical,gambetta2001state,negretti2013estimation,gammelmark2013bayesian,six2015parameter}. The parameters to be estimated may be the initial state for state tomography, or the parameters of the dynamics entering in the Hamiltonian, jump operators, or measurement efficiencies. Given a set of $n_\text{exp}$ independent realizations of the experiment $Y=\{(I_1^{(j)},\dots,I_n^{(j)})\}_{1\leq j \leq n_\text{exp}}$, the objective is to find the parameters $\theta$ that maximize the log-likelihood of the observed data:
\begin{equation}
    l_Y(\theta)=\sum_{j=1}^{n_\text{exp}}\log\left(p^\theta_{\bI_1,\dots,\bI_n}(I_1^{(j)},\dots,I_n^{(j)})\right),
\end{equation}
where $p^\theta_{\bI_1,\dots,\bI_n}$ is the probability density of the digitized signal for a fixed value of $\theta$. The latter is typically estimated using one of the previously discussed discretization schemes, with an approximate map $\tilde\Kcal_{I_k}$:
\begin{equation}
    p^\theta_{\bI_1,\dots,\bI_n}(I_1,\dots,I_n) = \mathrm{Tr}\big[\tilde\Kcal^\theta_{I_n}\dots \tilde\Kcal^\theta_{I_1}(\rho^\theta_0)\big].
\end{equation}
Since the digitization time is not infinitesimally small, this approximation introduces a constant bias in the estimate. In contrast, replacing the approximate map $\tilde\Kcal_{I_k}$ with the exact map $\Kcal_{I_k}$ derived in this paper yields an unbiased estimate, for any digitization time $\Dt$. This is illustrated for the frequency estimation of a qubit in \cref{fig:maxlike}.

Furthermore, in this example we observe that the variance of the correct log-likelihood computed with \texttt{Robinet} begins to saturate at $\Dt=0.1$. Thus, for the parameter to be estimated, most of the information in the continuous signal is already captured by the digitized signal at this timescale. Interestingly, for existing discretization schemes, the timescale $\Dt$ needed to obtain a sufficiently unbiased estimator can be orders of magnitude smaller! Again, robinet allows to discretize at the timescale that is informationally relevant, without worrying about the bias or stability of the reconstruction.

Finally, our method can also be used to sample trajectories for a finite $\Dt$, i.e., directly simulating the signal $I_k$ and the state $\robinet_k$ at each time step. To sample $I_k$ with the correct statistics, we use the probability distribution of the signal \cref{eq:Kcal-prob} with a rejection-sampling algorithm. For small $\Dt$, we can use the perturbative expansion \cref{eq:Kcal_I_p} to get the probability density $p_{\bI_k}$ of the random variable $\bI_k$:
\begin{equation}
    p_{\bI_k}(I) = \Tr{\Kcal_{I}(\robinet_{k-1})}=\frac{e^{-\Irm^2/2}}{\sqrt{2\pi\Dt}} \sum_{j=0}^\infty \alpha_j \Irm^j,
\end{equation}
where $\alpha_j$ is the trace of the operators of the corresponding order $\Irm^j$ in \cref{eq:Kcal_4} (and thus of order at least $\Dt^{j/2}$). So we can just precompute the coefficients $\alpha_j$ up to some order, and sample from the resulting probability distribution, a polynomial times a Gaussian, which can be done efficiently. This gives us a sample $I_k$ which we then use to get the next $\robinet_k$, and we repeat this process to sample a complete quantum trajectory.

\noindent\textbf{Conclusion} -- The robinet state that we have introduced is theoretically natural: it corresponds to the best state given the available information. Perhaps surprisingly, it can be computed exactly and efficiently numerically, enabling a wide range of computations at finite $\Dt$ without losing exactness. Compared to previous methods, it already allows for better state reconstruction and parameter estimation with existing experimental data. Furthermore, it eliminates the need for the very fine temporal binning that is usually required by existing schemes to ensure stable state reconstruction or unbiased estimation. Consequently, it opens the possibility of storing only coarsely-grained measurement signals sampled at the informationally relevant timescale; a new freedom that we hope experimentalists will consider in the future.

\begin{acknowledgments}
\noindent\textbf{Acknowledgments} --
We thank Francesco Albarelli, Felix Binder, Areeya Chantasri, Raphael Chetrite, Ronan Gautier, Marco Genoni, Hector Hutin, Marco Radaelli, and Howard Wiseman for discussions. For the numerics in this paper, we used the Dynamiqs library \cite{guilmin2025dynamiqs} designed for high-performance quantum systems simulation with JAX \cite{jax2018github} and Diffrax \cite{kidger2022on}. This project has received funding from the Plan France 2030 through the project ANR-22-PETQ-0006, as well as from the European Research Council (ERC) under the European Union’s Horizon 2020 research and innovation programme (grant agreement No. 884762), and also from the ERC QFT.zip project (grant agreement No. 101040260). This research was supported in part by the International Centre for Theoretical Sciences (ICTS) for participating in the program Quantum Trajectories (code: ICTS/QuTr2025/01).
\end{acknowledgments}
\bibliography{bib}

\pagebreak
\appendix
\section{Generalization}
The general formula for the jump and/or diffusive monitoring of $m$ jump operators $(L_1, \dots, L_m)$ with measurement record $J_k=(I_k^{(1)}, \dots, I_k^{(m)})$ at step $k$ is
\begin{equation}
    \Kcal_{J_k} = \left(\frac{1}{2\pi}\right)^{m} \!\!\!\int\!\!\dots\!\!\int \ddp_1\dots\ddp_m\,
    e^{i \sum_d p_d I_k^{(d)}} e^{\Dt\,\Lcal^{p_1,\dots,p_m}},
\end{equation}
where for each $p_d$ the bounds of the integral are from $-\pi$ to $\pi$ for a jump measurement and from $-\infty$ to $\infty$ for a diffusive measurement, and with $\Lcal^{p_1,\dots,p_m}$ the generalized tilted Liouvillian defined by
\begin{equation}
    \begin{split}
        \Lcal^{p_1,\dots,p_m} = \Lcal &- \sum_{d\in S_\text{diff}} \left(ip_d\Ccal_{L_d} + \frac{p_d^2}{2}\Ical\right)\\
        &+ \sum_{d\in S_\text{jump}}\left(e^{-ip_d} - 1\right)\Ccal_{L_d}.
    \end{split}
\end{equation}

\section{Perturbative expansion}
We derive a perturbative expansion of the completely positive map defined in \cref{eq:Kcal} to an arbitrary order in $\sDt$. For a fixed value of the digitized signal $I$, and for a time-independent Liouvillian, we recall the expression of $\Kcal_I$:
\begin{equation}
    \Kcal_I = \frac{1}{2\pi} \int_{\mathbb{R}} \ddp\, e^{i pI- \Dt\frac{p^2}{2}} e^{\Dt(\Lcal -ip \Ccal)}.
\end{equation}
As in the main text, we use the reduced variable ${\Irm=I/\sDt}$ of order $1$, and we make the change of variable ${x=p\sDt}$:
\begin{align}\label{eq:Kcal_x}
    \Kcal_I &= \frac{1}{2\pi\sDt} \int_{\mathbb{R}} \dx\, e^{-x^2/2+i \Irm x} e^{\Dt\Lcal -i x\sDt\,\Ccal}.
\end{align}
The strategy is to (i) Taylor expand the second exponential, (ii) integrate the different powers $x^k$ exactly against the Gaussian kernel $e^{-x^2/2+i \Irm x}$, and (iii) regroup the terms by their order in $\sDt$. We start by using the Taylor expansion
\begin{equation}\label{eq:taylor}
    e^{A+B} = \sum_{n=0}^\infty\frac{1}{n!}(A+B)^n = \sum_{n=0}^\infty\frac{1}{n!}\sum_{k=0}^n\lbb A^{n-k}B^k\rbb,
\end{equation}
where $\lbb A^{n-k} B^k\rbb$ is defined in the main text. We use this expression to expand $e^{\Dt\Lcal -i x\sDt\,\Ccal}$ in \cref{eq:Kcal_x} and obtain
\begin{equation}
    \begin{split}
        \Kcal_I =&~ \frac{1}{2\pi\sDt} \sum_{n=0}^\infty\frac{1}{n!} \sum_{k=0}^n\int_{\mathbb{R}}\dx\\
        &\, e^{-x^2/2+i \Irm x}(-ix)^k\sDt^{2n-k}\lbb\Lcal^{n-k} \Ccal^k\rbb.
    \end{split}
\end{equation}
We now integrate each power $x^k$ using the integral representation of the probabilist Hermite polynomial $\Hrm_k$ of order $k$:
\begin{equation}\label{eq:hermite_identity}
    \frac{1}{\sqrt{2\pi}}\int_{\mathbb{R}}\dx\, e^{-x^2/2+i \Irm x}(-ix)^k = e^{-\Irm^2/2}\Hrm_k(\Irm),
\end{equation}
which gives
\begin{equation}
    \Kcal_I = \frac{e^{-\Irm^2/2}}{\sqrt{2\pi\Dt}} \sum_{n=0}^\infty\frac{1}{n!}\sum_{k=0}^n \Hrm_k(\Irm)\sDt^{2n-k}\lbb\Lcal^{n-k} \Ccal^k\rbb.
\end{equation}
By changing the sum indices to group all terms with the same order $q=2n-k$ in $\sDt$, we obtain the perturbative expansion to any order $q$:
\begin{equation}\label{eq:Kcal_p}
    \Kcal_I = \frac{e^{-\Irm^2/2}}{\sqrt{2\pi\Dt}} \sum_{q=0}^\infty \sDt^q \sum_{n=\lceil q/2\rceil}^{q}\frac{\Hrm_{2n-q}(\Irm)}{n!}\lbb\Lcal^{q-n} \Ccal^{2n-q}\rbb,
\end{equation}
which is exactly \cref{eq:Kcal_I_p} in the main text.

Numerically, instead of using this exponentially growing formula, one can recursively compute each term in the Taylor expansion of $e^{\Dt \Lcal-ix\sDt \Ccal}$ in \cref{eq:Kcal_x} applied to $\rho$, up to the desired order in $\sDt$. This involves tracking terms of different orders in $x$ and then performing the exact integration using \cref{eq:hermite_identity}. The cost is thus quadratic in the chosen order $q$, as we apply $q$ times the superoperators to $q$ different terms.

For perfect detection efficiency $\eta=1$, the map can be written up to order five included with a single Kraus operator: ${\Kcal_I(\rho)=M_I\rho M_I^\dag +\mathcal{O}(\sDt^6)}$. Introducing ${G=-iH-\frac12 L^\dag L}$ and ignoring the Gaussian factor, the Kraus operator $M_I$ is defined by
\begin{align}
    M_I &\propto I \notag\\
    &+ \sDt^1\Irm\,L \notag\\
    &+ \sDt^2\left(G+\frac{\Irm^2-1}{2}L^2\right) \\
    &+ \sDt^3\left(\frac{\Irm}{2}\lbb GL\rbb + \frac{\Irm^3-3\Irm}{3!}L^3\right) \notag\\
    &+ \sDt^4\left(\frac{1}{2} G^2+\frac{\Irm^2-1}{3!}\lbb GL^2\rbb+\frac{\Irm^4-6\Irm^2+3}{4!}L^4\right) \notag\\
    &+ \sDt^5\left(
        \frac{\Irm}{3!}\lbb G^2L\rbb+\frac{\Irm^3-3\Irm}{4!}\lbb GL^3 \rbb\right. \notag\\
    & \qquad\qquad\qquad\qquad\qquad\ \ +\left. \frac{\Irm^5-10\Irm^3+15\Irm}{5!}L^5
    \right). \notag
\end{align}

\end{document}